\documentclass[prl,superscriptaddress,twocolumn,nopreprintnumbers,
floatfix,preprintnumbers,altaffilletter]{revtex4}
\usepackage{graphicx,url,amssymb,amsmath,longtable,rotating,color,units,
wasysym,subfigure,epsfig,multirow}

\begin{document}

\title{Detectability of Gravitational Waves from High-Redshift Binaries}

\author{Pablo A. Rosado}
\affiliation{Centre for Astrophysics \& Supercomputing, Swinburne University of Technology, PO Box 218, Hawthorn VIC 3122, Australia}
 
\author{Paul D. Lasky}
\author{Eric Thrane}
\affiliation{Monash Centre for Astrophysics, School of Physics and Astronomy, Monash University, VIC 3800, Australia}

\author{Xingjiang Zhu}
\affiliation{School of Physics, University of Western Australia, Crawley WA 6009, Australia}

\author{Ilya Mandel}
\author{Alberto Sesana}
\affiliation{School of Physics and Astronomy, University of Birmingham, Birmingham B15 2TT, UK}

\begin{abstract}
Recent non-detection of gravitational-wave backgrounds from pulsar timing arrays casts further uncertainty on the evolution of supermassive black hole binaries.
We study the capabilities of current gravitational-wave observatories to detect individual binaries and demonstrate that, contrary to conventional wisdom, some are in principle detectable throughout the Universe.
In particular, a binary with rest-frame mass $\gtrsim10^{10}\,M_\odot$ can be detected by current timing arrays at arbitrarily high redshifts.
The same claim will apply for less massive binaries with more sensitive future arrays.
As a consequence, future searches for nanohertz gravitational waves could be expanded to target evolving high-redshift binaries.
We calculate the maximum distance at which binaries can be observed with pulsar timing arrays and other detectors, properly accounting for redshift and using realistic binary waveforms.
\end{abstract}

\maketitle

{\it Introduction.}---
Gravitational-wave (GW) observatories are reaching remarkable sensitivities.
Advanced LIGO (aLIGO)~\citep{LIGO2015} and Virgo~\citep{AcerneseEtAl2015} are entering a new era, where the detection of GWs from compact binary mergers is expected to become a regular occurrence \citep{LIGO2010}.
At lower frequencies, Pulsar Timing Arrays (PTAs)~\citep{KramerEtAl13,ManchesterEtAl2013,MclaughlinEtAl13} are improving their sensitivity to supermassive black hole binaries (SBHBs)~\citep[e.g.,][]{Sesana2013B}.
The lack of a SBHB stochastic background detection might stem from the mechanism driving them to merge in cores of galaxies~\citep{ShannonEtAl2015}.
In particular, accelerated orbital evolution due to efficient coupling with the environment results in a sparser distribution of SBHBs~\citep{KocsisSesana2011}.
Whether in such circumstances a background is still more likely to be detected than individual binaries~\citep{SesanaVecchio2010, Sesana2013, RosadoEtAl2015} has not yet been investigated.

The conventional wisdom is that the most distant GW sources detectable by current PTAs are chirp mass $M_c\gtrsim10^{10}\,M_\odot$ binaries with redshifts $z\approx0.2$~\citep[e.g.,][]{ZhuEtAl2014,BabakEtAl2015,ArzoumanianEtAl2014} where $M_c$ is defined below.
However, this is false because of an effect we call {\it redshift bias}; actually, current PTAs are sensitive to some SBHBs at arbitrarily high $z$.

\begin{figure}[ht]
\begin{center}
\includegraphics{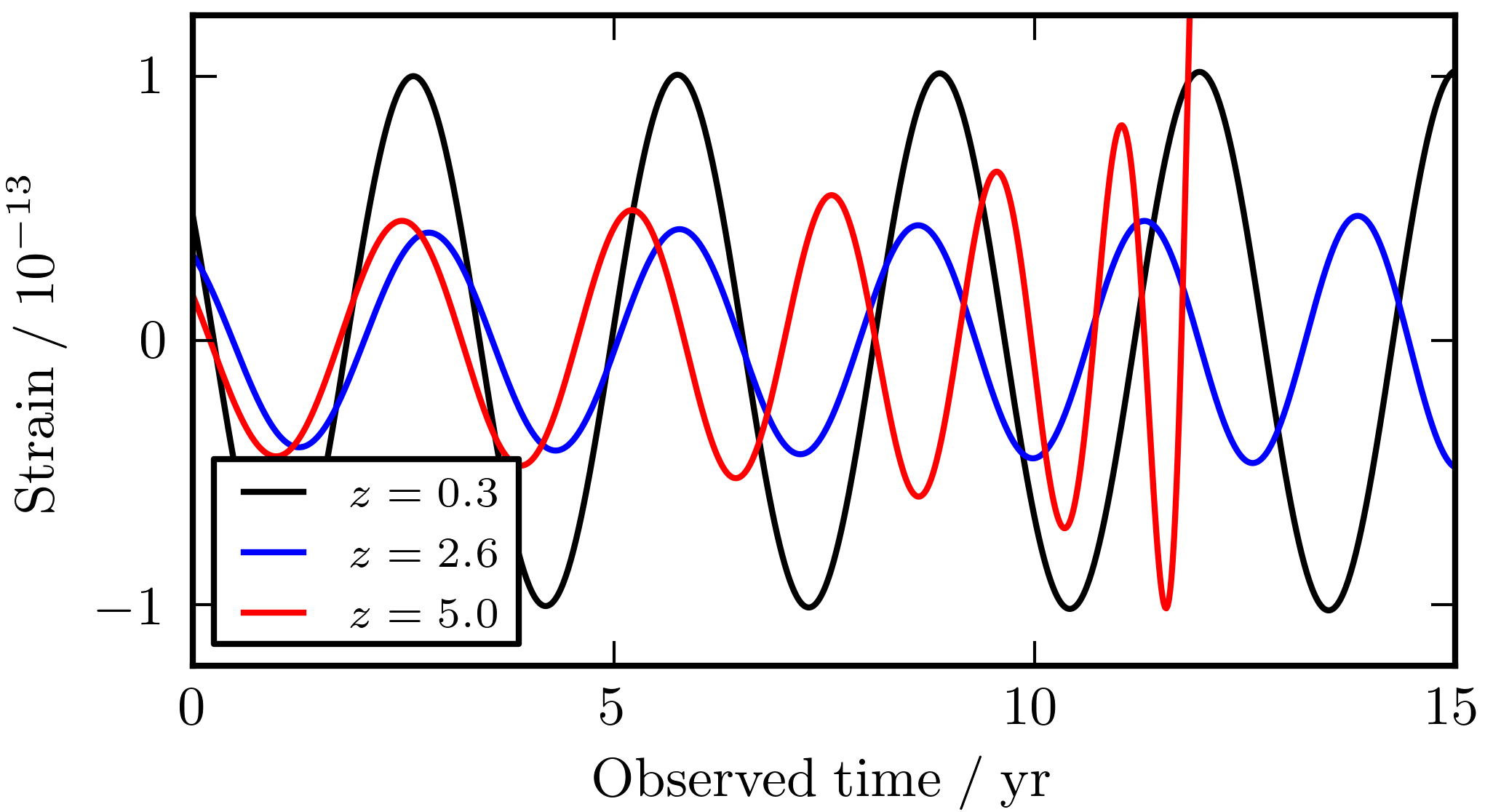}
\caption{\label{fg:waveforms} GW strain versus observed time for binaries with fixed rest-frame chirp mass, emitting at the same observed frequency at time $0$.
For clarity, we plot inspiral-only waveforms.
The higher the redshift, the closer the signal is to merger, causing the binary to be brighter than a similar, less distant one.
}
\end{center}
\end{figure}

Qualitatively, redshift bias works as follows.
During binary inspiral, the {\it emitted} frequency and strain amplitude increase with time.
Cosmological redshift causes the {\it observed} frequency to decrease with distance.
For some detector with fixed observed frequency, a high-$z$ system can produce a higher strain than a low-$z$ one with the same rest-frame mass because the former was at a later stage of the inspiral, emitting brighter GWs.
Consequently, sensitivity to binaries does not always decrease monotonically with $z$, but can instead have a minimum at a redshift $z_\text{min}$.
This can be counterintuitive: a binary at $z>z_\text{min}$ may look brighter than another with same mass at $z_\text{min}$ because, if both are observed in the same frequency band, the $z>z_\text{min}$ binary is intrinsically brighter despite being more distant.
Hence, by observing a binary in a specific band, there is a bias in which a different stage of the inspiral is selected.
We illustrate this in Fig.~\ref{fg:waveforms}.

The described effect is a GW analogue to the increase of apparent brightness with redshift of some known astronomical objects, accounted for with a negative K-correction~\citep[e.g.,][]{Hubble1936, OkeSandage1968}.
While redshift bias has been considered in PTA rate calculations~\citep{WenEtAl2011}, its consequences have not been fully investigated until now.
We quantify the effect of redshift bias and calculate the maximum redshift at which binaries are detectable, using gravitational waveforms that include merger and ringdown phases.
To be clear, we are not predicting an imminent detection of high-$z$ binaries.
The rate of binaries with appropriate masses, redshifts, and frequencies is unknown and could be small.
Nonetheless, given the current lack of SBHB observations, it is important to correctly describe the reach of extant and future PTAs.

\newpage

{\it Redshift bias.}---
The GW strain amplitude of an inspiral with component masses $m_1$, $m_2$ is~\citep{Maggiore2008}
\begin{equation}\label{eq:h}
h(t) = \frac{4\pi^{2/3}\left[GM_c\left(1+z\right)\right]^{5/3}}{c^4 D_L(z)}\left[f(t)\right]^{2/3},
\end{equation}
where $G$ is Newton's constant, $c$ is the speed of light, 
$M_c = (m_1m_2)^{3/5}/(m_1+m_2)^{1/5}$
is the rest-frame chirp mass, and $D_L(z)$ is luminosity distance.
Invoking the stationary phase approximation~\cite[e.g.,][]{DrozEtAl1999}, the Fourier transform of $h(t)$ is
\begin{equation}\label{eq:tildeh}
\tilde{h}(f)=\sqrt{\frac{5}{24}}\frac{\left[GM_c(1+z)\right]^{5/6}}{\pi^{2/3}c^{3/2}D_L(z)}f^{-7/6}.
\end{equation}
The observed GW frequency, $f(t)$, evolves as
\begin{equation}\label{eq:fdot}
\frac{df}{dt} = \frac{96\pi^{8/3}\left[G M_c (1+z)\right]^{5/3}}{5c^5}\left[f(t)\right]^{11/3},
\end{equation}
where we assume circular orbits.
Equations~(\ref{eq:h})--(\ref{eq:fdot}) imply that waveforms are determined by the commonly named ``redshifted chirp mass'': $M_z = (1+z) M_c$.

Redshifted chirp mass was introduced in~\cite{KrolakSchutz1987}, which showed that $M_c$ and $z$ are degenerate variables in all GW observables.
By introducing $M_z$, the redshift conveniently vanishes from the equations.
It is therefore widely used in GW literature, often without distinction from $M_c$.
The degeneracy implies that GW observations alone cannot produce a Hubble diagram, although, see \citep[e.g., ][]{TaylorEtAl2012, MessengerRead2012,Schutz1986, DelPozzo2012}.
A cosmological model can however break the degeneracy if $df/dt>0$ is measured.

GW detector sensitivity is customarily characterized by the ``horizon distance'': the maximum distance where binaries are detectable.
We assume a binary is detectable if its optimal signal-to-noise ratio
\begin{equation}\label{eq:snr}
{\rm S/N}=\left(  4 \int_{f_\text{min}}^{f_\text{max}} \frac{|\tilde{h}(f)|^2}{S_n(f)} df \right)^{1/2}
\end{equation}
exceeds a conservative threshold of $8$, which ensures a detection probability $>95\%$ for a false alarm probability $<0.1\%$ (in the context of single-source detection statistics, e.g.,~\cite{RosadoEtAl2015}).
In Equation~(\ref{eq:snr}), $S_n(f)$ is the detector strain noise, and $f_\text{min}$ is the minimum (i.e. initial) frequency at which the binary is observed.
If the binary does not merge during the observation, the maximum frequency $f_\text{max}$ is obtained by integrating Eq.~(\ref{eq:fdot}) over the observation time.
If the binary merges, $f_\text{max}$ is set sufficiently high so that the contribution to S/N from frequencies $>f_\text{max}$ is negligible.

The strain amplitude of an inspiral of fixed $M_c$ and $f$ [Eq.~(\ref{eq:h})] decreases with $z$ until $z_\text{min}$ and then increases.
The redshift $z_\text{min}$, calculated by solving $\partial h(t)/\partial z=0$, satisfies
\begin{equation}\label{eq:snrmin}
(1+z_\text{min})\frac{d\ln[D_L(z)]}{dz}\bigg|_{z_\text{min}}=\frac{5}{3}.
\end{equation}
This result is independent of the detector and assumed cosmology, and is also independent of the binary characteristics provided its rest-frame frequency $f(1+z)$ is in the inspiral phase.
Adopting standard $\Lambda$CDM cosmology~\citep{PlanckCollaboration2015}, $z_\text{min}\approx 2.63$.

\begin{figure}[h]
\begin{center}
\includegraphics{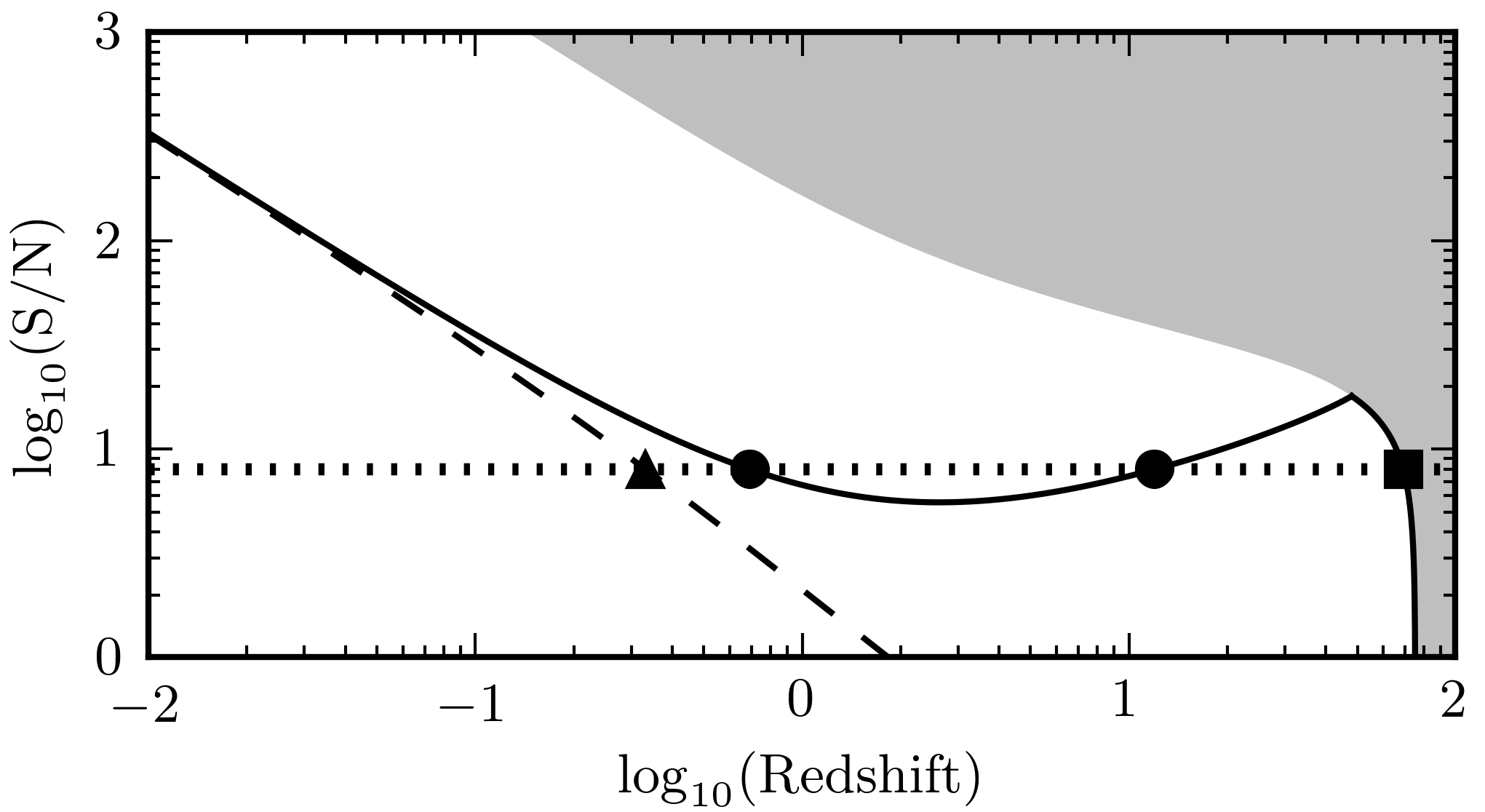}
\caption{\label{fg:snr} S/N versus redshift of inspiralling binaries of certain mass and GW initial observed frequency, and certain detector noise.
The dotted line is the S/N threshold and the grey area contains the parameter space where binaries already merged.
The solid and dashed lines are respectively the S/N with fixed $M_c$ and $M_z$.
The triangle and square denote the horizon distance inferred given fixed $M_z$ and $M_c$ respectively.
The two circles define a region where the binary S/N is below the threshold.}
\end{center}
\end{figure}

In Eq.~(\ref{eq:snrmin}), $z_\text{min}$ is also the S/N minimum assuming $S_n(f)$ is approximately constant between $f_\text{min}$ and $f_\text{max}$, which holds if the binary evolution is slow enough during the observation time.
This is not fulfilled by ground-based detectors, where all binaries observed $\gtrsim1\,$Hz coalesce in less than a typical observation time.
However, it does hold for lower-frequency detectors, such as PTAs and future observatories like eLISA \citep{eLISA2013}.
This has an interesting implication for such GW observatories: an instrument capable of detecting slowly inspiralling binaries of certain mass from a particular initial frequency and redshift $z_\text{min}$, can detect binaries of the same mass and frequency at much higher redshifts.
This is illustrated in Fig.~\ref{fg:snr}.

\begin{figure*}[ht]
\begin{center}
\includegraphics{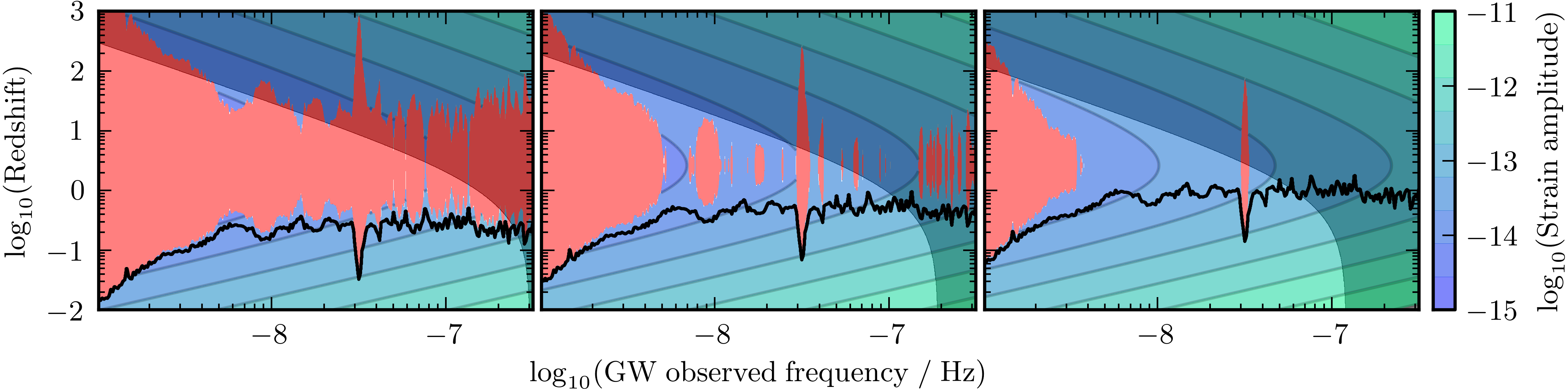}
\caption{\label{fg:EPTA} Redshift versus GW observed frequency of SBHBs, assuming they are monochromatic.
The blue-green area shows where SBHBs with $M_c=10^{9.8} M_\odot$ (left), $10^{10.0} M_\odot$ (middle), and $10^{10.2} M_\odot$ (right) produce a strain larger than the EPTA upper limit~\cite{BabakEtAl2015}.
In the shaded area binaries already finished the inspiral phase, and the red area contains SBHBs with sub-threshold strains.
The black line shows the horizon distance from~\citep{BabakEtAl2015} assuming $M_z$ (not $M_c$) is $10^{9.8} M_\odot$ (left), $10^{10.0} M_\odot$ (middle) and $10^{10.2} M_\odot$ (right).
}
\end{center}
\end{figure*}

Figure~\ref{fg:snr} shows the dependence of S/N with $z$ for fixed $M_c$ (solid line) and $M_z$ (dashed).
In the former case, redshift bias causes the turnover.
In the latter case, fixing $M_z$ implies the rest-frame chirp mass decreases as $z$ increases, implying the S/N decreases monotonically.
Figure~\ref{fg:snr} depicts three interesting implications of redshift bias.
First, identifying $M_z=M_c$ is inaccurate: one would claim that binaries of such mass cannot be seen beyond $z\sim 0.3$ (triangle), whereas in principle they are observable to $z\sim 70$ (square).
Second, there can be a $z$-interval (between the circles) that is undetectable.
Finally, for sufficiently large $M_c$, the solid curve exceeds the threshold at all $z$.
{\it Thus, binaries with such $M_c$ and observed frequency can be detected at any distance.}
And, as detector sensitivity improves, the value of $M_c$ for which the horizon distance extends to arbitrarily high $z$ drops.

{\it Horizon distances with current PTAs}.---
Under the common assumption that SBHBs are approximately monochromatic, S/N and strain thresholds are interchangeable.
The EPTA~\citep{KramerChampion2013} has the most stringent upper limits on individual SBHB strains~\citep{BabakEtAl2015}.
Figure~9 of that paper shows horizon distances for different values of $M_z$.
Figure~\ref{fg:EPTA} of this paper shows those same horizon distances as black lines, assuming fixed $M_z$.
For simplicity, we assume $m_1=m_2$ hereafter.

The plots in Fig.~\ref{fg:EPTA} show the maximum strain with different fixed values of $M_c$.
These values ($10^{9.8} M_\odot$, $10^{10.0} M_\odot$, and $10^{10.2} M_\odot$) are chosen for illustrative purposes: for $M_c>10^{10.2} M_\odot$, all binaries observed at $\gtrsim 10^{-9}$\,Hz are detectable, whereas for $M_c<10^{9.8} M_\odot$ redshift corrections become small.
The turnover at $z_\text{min}\approx 2.63$ produced by redshift bias is noticeable.
Shaded areas are where rest-frame frequencies exceed the innermost stable circular orbit (ISCO); a proxy for the end of the inspiral phase.
Red areas account for binaries whose strain is below the EPTA upper limit.
Blue-green areas contain binaries with fixed $M_c$ and strain above the upper limit; binaries here are considered detectable, assuming they are monochromatic.
When $M_c\gtrsim 10^{10}\,M_\odot$, the EPTA becomes sensitive to binaries beyond $z_\text{min}$ in a wide frequency band, and hence anywhere in the Universe.
Similar plots can be obtained for the PPTA~\cite{ZhuEtAl2014} and NANOGrav~\cite{ArzoumanianEtAl2014}.

While Fig.~\ref{fg:EPTA} illustrates redshift bias applied to real data, the assumption of monochromatic SBHBs is not generally true, especially for high-$z$ binaries that are closer to coalescence.
We now study detectability more thoroughly by imposing S/N$>8$ and taking into account binary evolution and finite observation times, which enter the calculations through $f_\text{max}$ in Eq.~(\ref{eq:snr}).
Furthermore, we use approximate (non-spinning, quasi-circular) inspiral-merger-ringdown waveforms~\citep{AjithEtAl2008}.
For simplicity, previous explanations focused only on the inspiral phase; similar arguments hold during merger and ringdown.

\begin{figure*}[ht]
\begin{center}
\includegraphics{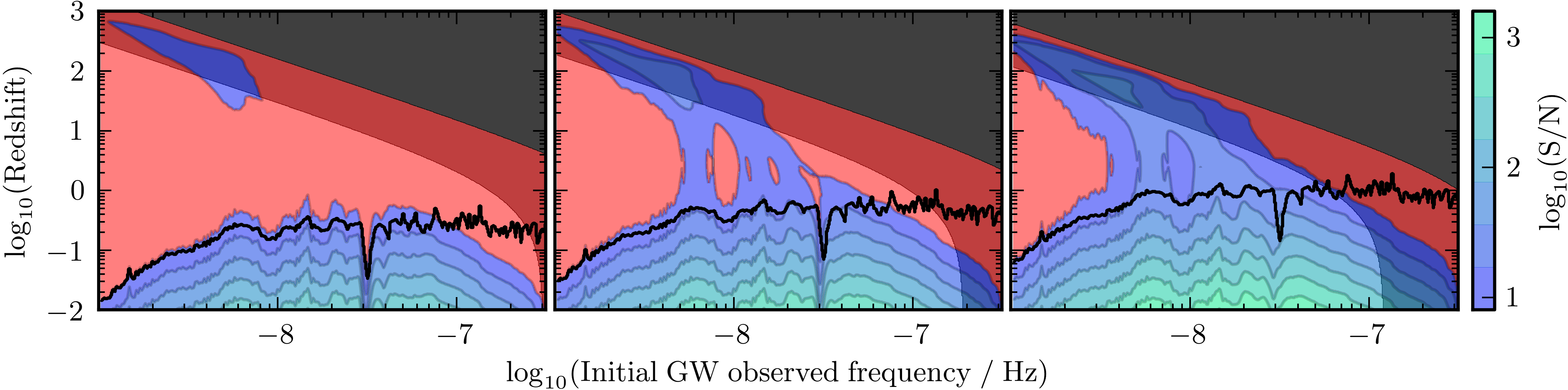}
\caption{\label{fg:EPTA2} Redshift versus initial GW observed frequency ($f_\text{min}$) at which binaries with $M_c=10^{9.8} M_\odot$ (left), $10^{10.0} M_\odot$ (middle), and $10^{10.2} M_\odot$ (right) are detectable, assuming 10-year observations.
Red and dark grey areas contain undetectable or merged binaries, respectively.
The color of blue-green areas gives the S/N of detectable binaries consistent with the EPTA upper limit~\cite{BabakEtAl2015}.
Light-shaded areas contain binaries in the merger and ringdown phases.
}
\end{center}
\end{figure*}

Figure~\ref{fg:EPTA2} shows $z$ versus $f_\text{min}$ (initial observed frequency), assuming 10-year observations.
The color scale now gives S/N.
The noise $S_n(f)$ is chosen such that monochromatic binaries, producing a strain equal to the EPTA upper limit at frequency $f$, would have S/N$=8$.
This is rather optimistic: the S/N threshold corresponding to the upper limit should be smaller; however, with this choice the effects of binary evolution and the inclusion of merger and ringdown phases can be better appreciated by direct comparison of Figs.~\ref{fg:EPTA} and~\ref{fg:EPTA2}.
The region of detectable binaries differs significantly from that depicted in Fig.~\ref{fg:EPTA}, where binaries are assumed to be monochromatic.
The light-shaded area contains binaries in the merger and ringdown phases at the start of the observation.
In the dark-shaded area binaries already merged.

\begin{figure*}
\begin{center}
\includegraphics{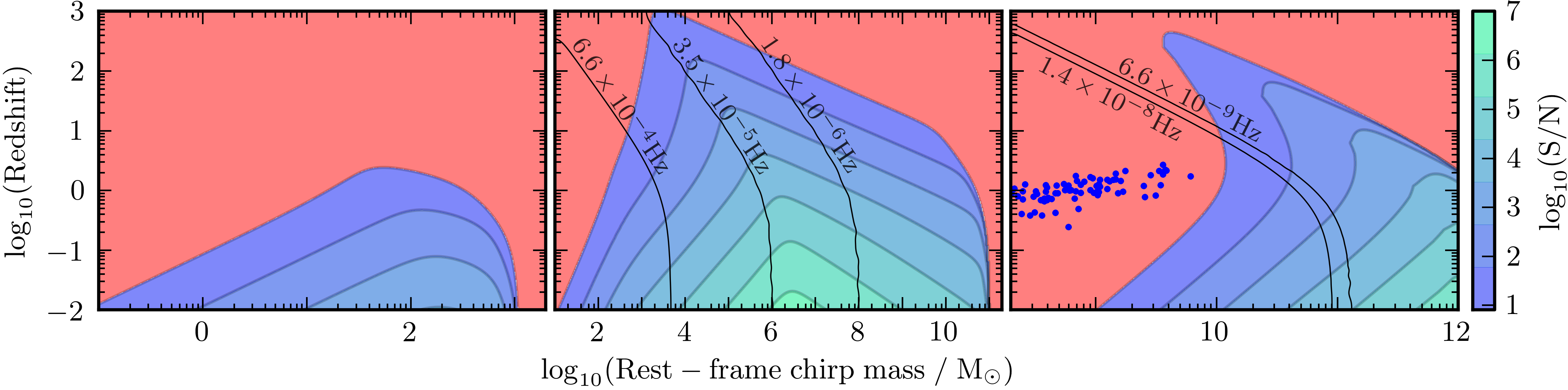}
\caption{\label{fg:z_vs_mc}
Horizon distances for aLIGO (left), LISA (middle), and PPTA (right).
The assumed observing time is 5 years for LISA and 10 for PPTA.
For aLIGO the observing time is irrelevant given the short duration of binaries in band.
Blue-green colors give the maximum S/N, whereas red areas contain undetectable binaries.
Thin black lines are contours of optimal $f_\text{min}$.
Blue dots are the binary candidates of~\cite{GrahamEtAl2015}.
}
\end{center}
\end{figure*}

{\it Conveying sensitivity to binaries.}---
Figure~\ref{fg:z_vs_mc} shows the detectable parameter space in terms of physical quantities, $z$ and $M_c$, for aLIGO (assuming the zero-detuning, high-power noise spectrum~\citep{LIGO2010b}), LISA (assuming the canonical design configuration, e.g.,~\cite{MooreEtAl2015}), and PPTA.
For the latter we calculate the combined S/N from the 20 pulsars in Table 3 of~\cite{ManchesterEtAl2013} using the noise models of~\cite{WangEtAl2015}.
Note that a binary cannot be optimally located and oriented with respect to all pulsars simultaneously; however, given that most of the S/N contribution usually comes from a few pulsars, one can consider that the binary location and orientation favors those pulsars.

The S/N depends on the frequency where a binary is initially observed, $f_\text{min}$.
We choose this frequency to give the highest S/N possible.
Black thin lines on the middle and right plots show contours of optimal $f_\text{min}$.
For aLIGO, this frequency always corresponds to the minimum frequency of the detector ($\lesssim 10\,$Hz).

The blue dots on the right-most plot of Fig.~\ref{fg:z_vs_mc} are recently-published SBHB candidates~\cite{GrahamEtAl2015}; we see that, even if these have optimal orientation and frequency, they are not detectable.
In fact, the maximum S/N among these candidates is $\sim 2$.

Horizon distances as a function of $M_c$ and $M_z$ are quantitatively different~\citep[e.g., Fig. 2 of][for aLIGO]{DwyerEtAl2015}.
This difference is particularly striking for PTAs: when plotting horizon distances versus $M_c$, one notices the S/N turnover caused by redshift bias.
The inclusion of merger and ringdown phases improves aLIGO horizons significantly~\citep{LIGO2013c}, since all binaries entering the band merge within a short observing timescale.
For PTAs, including merger and ringdown does not affect the detectability of low-mass SBHBs since the merger does not occur in the observable band, but extends the detectable region towards higher values of $M_c$ and $z$.

{\it Conclusions.}---
Redshifted chirp mass $M_z$ is a convenient, and widely used quantity in GW calculations.
However, conflating it with the rest-frame chirp mass $M_c$ can lead to a significant underestimation of the distance reach of a detector.
We calculate horizon distances by properly accounting for redshift, and use realistic waveforms including merger and ringdown.
This way, horizon distances of Advanced LIGO, LISA, and especially PTAs are significantly improved.

In fact, current PTAs are capable of detecting massive objects further away than any other existing astronomical observatory.
Optimally oriented binaries in the observing band with $M_c\gtrsim 10^{10}\,M_\odot$ can be detected at any distance, corresponding to a detectable volume $\sim$4000 times larger than previously claimed in the literature.
Assessing the corresponding increase in the binary detection rate would require relying on uncertain theoretical models.
Under the standard hierarchical model of supermassive black hole formation, major mergers are less likely at higher $z$~\citep[e.g.,][]{SesanaEtAl2004, Barausse2012}.
Moreover, the lifetime of a binary decreases with $z$; e.g., a $M_c=10^{10}\,M_\odot$ binary at $f=6\times10^{-9}$\,Hz, and $z=0.2$ (the horizon distance claimed in previous work) spends $\sim2.3$ times longer in the PTA band than another binary with equal $M_c$ and $f$ but $z=1$ (which is much smaller than our claimed horizon distance, but astrophysically more conservative).
Nonetheless, the volume of the Universe with $z<1$ is $\sim 65$ times larger than with $z<0.2$.
Finally, as PTA sensitivity improves, binaries with $M_c<10^{10}\,M_\odot$ will also become detectable at any distance.
Hence, future searches for individual binaries could consider high-$z$, non-monochromatic systems.

The GW strain amplitude that an inspiralling binary produces in a detector does not decrease monotonically with redshift.
Instead, it reaches a minimum and then increases.
The turnover, first noticed in~\cite{WenEtAl2011}, is accounted for by what we call \textit{redshift bias}, illustrated with the examples above.
The redshift at which strain is minimum, $z_\text{min}\approx 2.63$, is independent of binary and detector properties.
For low-frequency detectors such as PTAs and LISA in which binaries evolve slowly, there is also a minimum in the signal-to-noise ratio (and hence detection probability).
This implies, if such a detector is capable of detecting binaries at $z_\textrm{min}$ inspiralling at a certain observed frequency, it can detect inspiralling binaries with the same mass and observed frequency at any redshift.

A previous study~\citep{RosadoEtAl2015} showed that a GW background is more likely to be detected than individual binaries, under the assumption of circular, GW-driven binaries.
If recent PTA non-detections favour a low-frequency signal turnover~\citep{TaylorEtAl2015}, the low-$z$ SBHB population might indeed be sparser than expected.
In this context, putative high-$z$ massive binaries (caught at higher intrinsic frequency, and therefore likely unaffected by environmental coupling) might be easier to detect, a possibility that deserves investigation.

Finally, future work should address both the probability of detecting SBHBs under realistic, astrophysically-motivated frameworks, and also the implications that detections of high-$z$ binaries could have when contrasting different cosmological models.

\begin{acknowledgements} 
PAR thanks C. Blake, N. Cornish, N. Creighton, C. Messenger, F. Ohme, and especially Y. Wang for their useful comments and suggestions.
PAR and PDL are supported by the Australian Research Council Discovery Project DP140102578.
XZ is supported by the Australian Research Council Discovery Project DP150102988.
AS is supported by the Royal Society.
E. T. is supported by ARC FT150100281.
\end{acknowledgements}

{\it Note added}.--Recently, the first detection of GWs by aLIGO has been announced from a binary black hole merger at $z\approx 0.09$ \citep{AbbottEtAl2016}. With the projected increase in the sensitivity of aLIGO over the coming years, expanding the detector reach to higher redshifts, cosmological effects will play an increasingly important role.

\bibliography{horizon}

\end{document}